\documentclass[usenatbib]{mn2e}

\usepackage{graphicx}

\title[Detection of superhumps in KR~Aur]{Detection of superhumps in the VY~Scl-type \\ nova-like variable KR~Aur}

\author[V. P. Kozhevnikov]{V. P. Kozhevnikov\thanks{E-mail:valerij.kozhevnikov@usu.ru}\\
Astronomical Observatory, Ural State University, Lenin Av. 51, Ekaterinburg, 620083, Russia}
 
\begin{document}

\date{Accepted. Received; in original form}

\pagerange{\pageref{firstpage}--\pageref{lastpage}} \pubyear{}

\maketitle

\label{firstpage}

\begin{abstract} 
We report on detection of negative superhumps in KR~Aur which is the representative member of the VY~Scl stars. The observations were obtained with the multi-channel photometer during 107 h. The analysis of the data clearly revealed brightness variations with a period of $3.771 (\pm0.005)$~h. This is 3.5 per cent less than $P_{orb}$, suggesting it is a negative superhump. Negative superhumps in VY~Scl stars are widely spread. The discovery of powerful soft X-rays from V751~Cyg demonstrates that VY~Scl stars may contain white dwarfs, at which nuclear burning of the accreted material occur. If this suspicion is correct, we then can suppose that the powerful radiation emerging from the white dwarf may cause the tilt of the accretion disk to the orbital plane, and its retrograde precession may produce negative superhumps in VY~Scl stars. 
\end{abstract}


\section{INTRODUCTION}
Cataclysmic variables (CVs) are interacting binaries that consist of a white dwarf primary accreting matter from a low mass secondary. In non-magnetic systems this matter forms a bright accretion disc. Many CVs of short orbital period have light-curves with prominent humps, which reveal periods slightly longer than $P_{orb}$. These are called "superhumps" since they are characteristic of the SU~UMa-type dwarf novae in superoutburst. Some bright nova-like variables and nova remnants show superhumps during the normal brightness state. These are called permanent superhumps. Their periods can either be a few per cent longer than $P_{orb}$ and they are called "positive superhumps", or they can be a few per cent shorter -- "negative superhumps". Typical full amplitudes of permanent superhumps are about 5--15 per cent, but they are highly variable and sometimes even disappear from the light-curve. The periods of the superhumps are unstable and usually show appreciable wobbling. The positive superhump is explained as the beat between the binary motion and the precession of an eccentric accretion disc in the apsidal plane. The negative superhump is explained as the beat between the orbital period and the nodal precession of the disc tilted to the orbital plane. A review of permanent superhumps is given by \citet{patterson93}. Some recent data about permanent superhumps are given by \citet{patterson02}.

The VY~Scl stars are CVs, the light-curves of which are characterized by occasional drops from steady high states into low states lasting up to several hundred days. VY~Scl stars often reveal positive and negative superhumps. The classical example is TT~Ari which shows alternation between positive and negative superhumps \citep{skillman98}. KR~Aur is a representative member of the VY~Scl stars. Its spectroscopic observations revealed that it is a close binary system, consisting of a white dwarf (0.7 M$_{\sun}$) and a red dwarf (0.48 M$_{\sun}$) with an orbital period of 3.907~h  \citep{shafter83}.  A historical light-curve of KR~Aur was compiled by \citet{liller80}. The brightness of KR~Aur is usually between 12--14~mag, with occasional decreases to 15.5~mag, but it drops up to 18~mag occasionally. In this star \citet{singh93} detected large amplitude quasi-periodic oscillations (QPOs) with periods in the range 500--800 s. There were also indications of a modulation with a period of about 4~h. This modulation might represent the orbital period or might be a kind of superhump. However, \citet{kato02} conducted photometric observations of KR~Aur and found no modulation with such a period. We decided to verify whether a photometric signal near the orbital period is present in KR~Aur, and the first observational night clearly showed such a modulation. In this paper we present results of our observations spanning a total duration of 107 h.

\section{OBSERVATIONS} \label{observations}

In observations of CVs we use a multi-channel photometer that allows us to make continuous brightness measurements of two stars and the sky background. PMTs have lesser quantum efficiency in comparison with CCDs and are not advantageous in observations of faint stars. However, in observations of relatively bright stars (12-13~mag, meter-class telescopes) multi-channel photometers with PMTs can attain the same or better accuracy under similar conditions \citep{abbott94}. In this we made sure when performing the noise analysis of our observations according to the TEP network \citep{kozhevnikoviz}. The TEP network was aimed to detect planetary transits in CM~Dra which is a star of 11~mag. Our accuracy was nearly the same as the accuracy of the other TEP network observations with CCDs although we used a smaller telescope located in unfavorable photometric conditions (see tables 1, 3 of \citealt{deeg98}). 

KR~Aur was observed in 2004 January and February during 13 nights using the 70-cm telescope at the Kourovka observatory of the Ural State University. A journal of the observations is given in Table~\ref{journal}. The program and comparison stars were observed through 16 arcsec diaphragms, and the sky background was observed through a 30 arcsec diaphragm. The comparison star is located at the direction NE and at a separation of 7.4 arcmin from KR~Aur. This comparison star was chosen among nearby stars such that its color was close to the color of KR~Aur. This star was also chosen only slightly brighter than KR~Aur because this minimized the influence of the variable sky background under non-photometric conditions. Data were collected at 8-s sampling times in white light (approximately 300--800~nm), employing a PC based data acquisition system. We used the CCD guiding system that enables precise centring of the two stars in the diaphragms to be maintained automatically. This improves the accuracy of brightness measurements and facilitates the acquisition of long continuous light-curves. The design of the photometer is described by \citet{kozhevnikoviz}.

\begin{table}
\caption{Journal of the observations}
\label{journal}
\begin{tabular}{@{}l c c}
\hline
\noalign{\smallskip}
date & HJD start & length \\
(UT) & (-2\,453\,000) & (hours) \\
\hline
2004 Jan 14   & 19.14751   & 6.4   \\
2004 Jan 15   & 20.08139   & 11.3   \\
2004 Jan 22   & 27.08817   & 8.7   \\
2004 Jan 23 & 28.08144   & 6.5 \\
2004 Jan 24   & 29.08618   & 10.7  \\
2004 Jan 29   & 34.12206   & 10.0  \\
2004 Jan 30   & 35.10234   & 6.9  \\
2004 Feb 05   & 41.13106   & 7.4  \\
2004 Feb 15   & 51.10879   & 7.8  \\
2004 Feb 16    & 52.10926   & 8.8  \\
2004 Feb 20    & 56.11692   & 8.2  \\
2004 Feb 23    & 59.13386   & 6.8  \\
2004 Feb 28    & 64.12625  & 7.3  \\
\hline
\end{tabular}
\end{table}

\begin{figure}
\includegraphics[width=84mm]{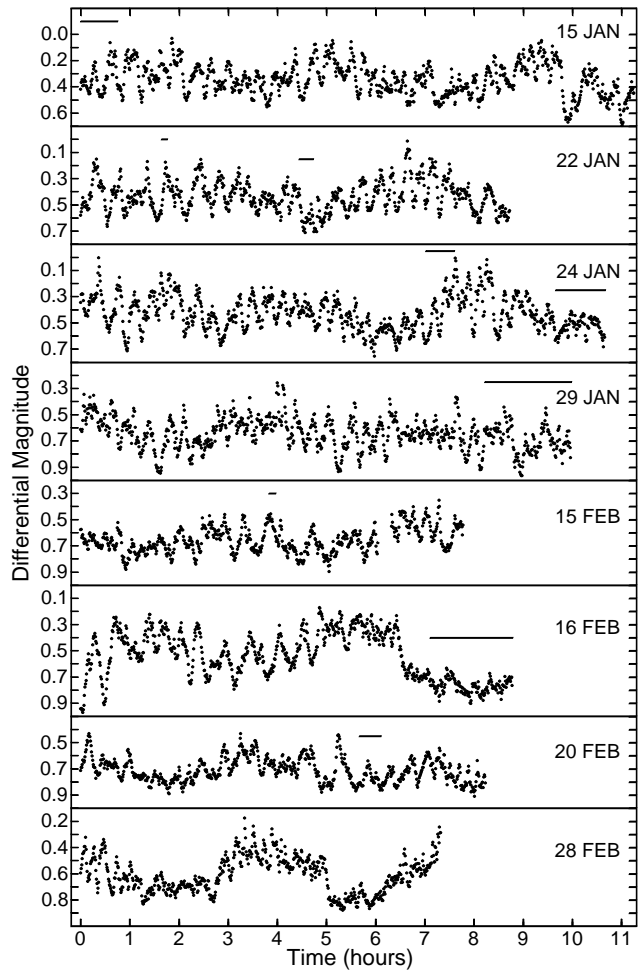}
\caption{Differential light-curves of KR~Aur. They show brightness variations with a period of about 4 h. The horizontal lines mark the time intervals when thin clouds have appeared. The nightly low-frequency trends are not removed.}

\label{krblkr}
\end{figure}

The measurements of the sky background were subtracted from the program and comparison star data, taking into account the differences in light sensitivity between the channels. We then took differences of magnitudes of the program and comparison stars. Because the angular separation between the program and comparison stars is small, the differential magnitudes are corrected for first order atmospheric extinction and light absorption by thin clouds that appeared sometimes during the observations. According to the mean counts, the photon noise (rms) of the differential light-curves equals 13~mmag. The actual rms noise also includes atmospheric scintillations and motion of the star images in the diaphragms. We estimate these noise components roughly equal to 5~mmag each. The total white noise component of the light-curves (rms) is then 15~mmag. Fig.~\ref{krblkr} presents the longest differential light-curves of KR~Aur with magnitudes averaged over 40-s time intervals. The white noise component of these light-curves is 7~mmag. Besides the white noise components each photometric system usually shows the $1/f$ noise component which decreases the precision at frequencies approximately below 1~mHz (e.g. \citealt{young91}). When observing in white light, most of the $1/f$ noise component can be caused by differential extinction. However, in the following analysis we use the light-curves, in which nightly low-frequency trends are removed by subtraction of a second order polynomial fit. After such a procedure the average level of the $1/f$ noise in the amplitude spectra at frequencies below 1~mHz is small and does not exceed 1~mmag. 

During the main observations we carried out the measurements only in white light. A month later we found $B$=13.47, $V$=13.35 for KR~Aur and $B$=13.26, $V$=12.70 for the comparison star. This implies that in white light KR~Aur was fainter than the comparison star approximately by 0.4~mag. As seen in Fig~\ref{krblkr}, the differential magnitude of KR~Aur was roughly the same. Hence the B magnitude of KR~Aur was about 13.5 during our observations. 

\section{ANALYSIS AND RESULTS}

As seen in Fig.~\ref{krblkr}, the light-curves of KR~Aur are fairly typical of CVs in showing rapid flickering. QPOs on a time-scale of tens of minutes are also easily visible. In addition, the light-curves show prominent maxima and minima, which may indicate periodic oscillation on a time-scale of hours. Fourier amplitude spectra calculated for longest de-trended light-curves show that the semi-amplitude of this oscillation is variable from night to night with the range 0.05--0.15~mag. However, since the oscillation has only a few oscillation cycles in each individual light-curve, these spectra cannot allow us to estimate the oscillation frequency with sufficient accuracy. When considering independently from each other, the individual light-curves also cannot show whether this oscillation is coherent. That is why we analysed the data incorporated into common time series. These time series were composed of all the de-trended individual light-curves and the gaps due to daylight and poor weather according to the observations. 

\begin{figure}
\includegraphics[width=84mm]{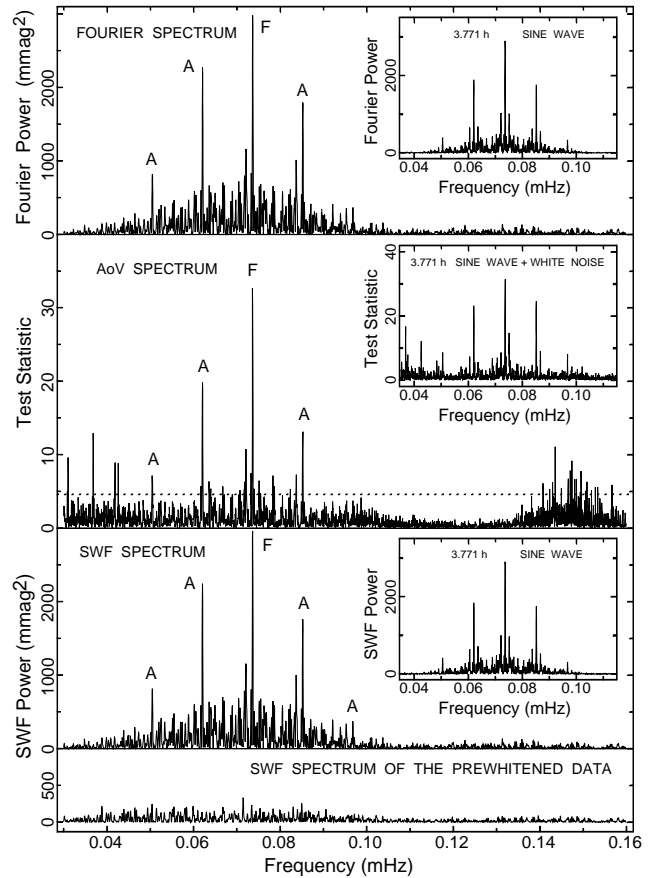}
\caption{Power spectra of KR~Aur that are calculated by means of a FFT algorithm, via the analysis of variance (AoV) of folded light-curves and with the aid of a sine wave fit (SWF) to folded light-curves. The horizontal dotted line marks the 0.1 per cent significance level.  The principal peaks are labelled with "F" and the one-day aliases are labelled with "A". The lower frame presents the SWF power spectrum of the prewhitened data.}
\label{swfpokr}
\end{figure}

To calculate power spectra, at first we used a fast Fourier transform (FFT) algorithm. The second method was the analysis of variance (AoV) \citep{schwarzenberg}. The AoV method usually applies to data folded and grouped into bins according to the phase of a trial period. But here we specify trial frequencies with a constant step of their change and then apply the AoV method as before. Then, instead of an AoV periodogram, we obtain an AoV spectrum, which can cover larger time intervals of variability. In addition, such an AoV spectrum is easily comparable with a Fourier power spectrum. However, the test statistic $\Theta_{\rm AoV}$ is sensitive also to oscillations with multiple frequencies, and this creates additional noise. That is why one more power spectrum was calculated with the aid of a sine wave fit (SWF) to folded light-curves, using the method of least squares. Such a power spectrum can have the large detection sensitivity of periodic signals that is inherent for AoV spectra. In addition, it is not sensitive to multiple frequencies. All these spectra are presented in Fig.~\ref{swfpokr}. They show distinct pictures of principal peaks and one-day aliases. The principal peaks correspond to a period of $3.771 (\pm0.005)$~h. The width and location in frequency of all the one-day aliases strictly correspond to the window functions obtained from artificial time series consisting of a sine wave and the gaps. This means that the observed oscillation behaves like strictly periodic oscillation and has coherent phase during all the observational nights. It leaves no doubt in the reality of the detected oscillation. 

As mentioned, we analysed de-trended individual light-curves. It is worthwhile to notice that the subtraction of a polynomial fit is a standard procedure used in time series analysis for de-trending  (e.g. \citealt{bendat}). Being able to only decrease amplitudes of signals at very low frequencies, this procedure cannot introduce additional periodicities. However, the oscillation observed in KR~Aur have a rather low frequency and, therefore, we must estimate the effect of this de-trending. Making numerical experiments with artificial time series, we found out that the subtraction of a second order polynomial fit from the individual light-curves decreases the amplitude of the 3.771-h signal only by 20 per cent. 

\begin{figure}
\includegraphics[width=84mm]{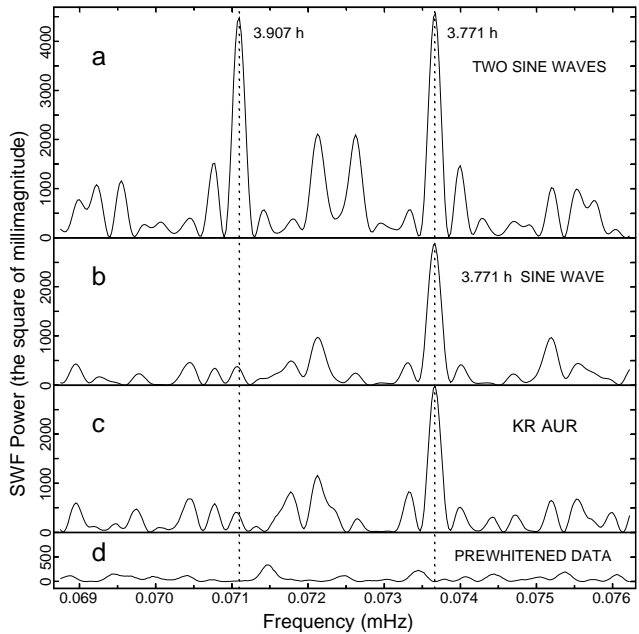}
\caption{Frame (a) presents the SWF power spectrum of an artificial time series consisting of two sine waves with periods of 3.907~h and 3.771~h and with the gaps. Frame (b) presents the SWF power spectrum of a sine wave with a period of 3.771~h and with the gaps. Frames (c) and (d) show the SWF power spectra of the data of KR~Aur and the prewhitened data in a large scale.}
\label{large}
\end{figure}

The period of the observed oscillation significantly differs from the orbital period of KR~Aur that equals 3.907~h \citep{shafter83}. The large noise level visible between the principal peak and the one-day aliases (Fig.~\ref{swfpokr}) may denote the presence of another oscillation with a similar frequency, and this oscillation may be the orbital period.  This signal can be hidden in relatively large features of the window function that are caused by the large 3.771-h signal. To find this oscillation, we excluded the main oscillation from the data. This is the well-known technique that allows us to remove these features and, therefore, makes visible the smaller signal in power spectra. Fig.~\ref{large} shows results. Fig.~\ref{large}(a) demonstrates the SWF power spectrum of an artificial time series where two large peaks indicate the frequencies of  two oscillations with periods of 3.907~h  and 3.771~h. This also shows that the frequency resolution of our data is sufficient to distinguish these two periodicities. Fig.~\ref{large}(b) shows the SWF power spectrum of an artificial time series consisting of a sine wave with a period of 3.771~h  and with the sampling of the observed data. This spectrum characterizes fine structure of the window function. The SWF power spectrum of KR~Aur is shown in Fig.~\ref{large}(c). This spectrum reveals a remarkable similarity with the SWF power spectrum presented in Fig.~\ref{large}(b). This means that most of the features in the power spectrum of KR~Aur correspond to the window function, and this, in turn, denotes the presence of a single periodic signal. Indeed, as seen in Fig.~\ref{large}(d), the subtraction of the sine wave with a period of 3.771~h and with the appropriate amplitude and phase removes most of these features. The complete absence of any feature coinciding with the orbital frequency especially impresses. Hence there is no detected photometric signal with $P = P_{orb}$. Other similar periods are also absent. The large range of the power spectrum of the prewhitened data is presented in the lower frame of Fig.~\ref{swfpokr}.

\begin{figure}
\includegraphics[width=84mm]{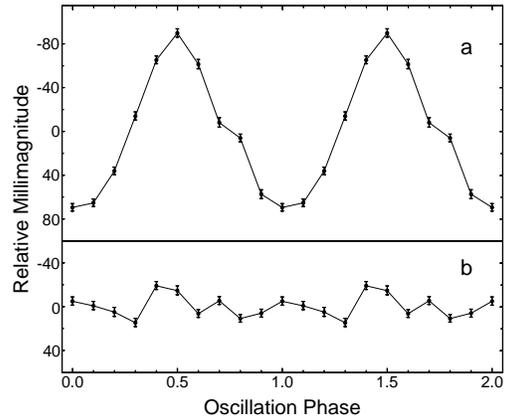}
\caption{Light-curves of KR~Aur that are folded with a period of 3.771~h (frame a) and with a period of 3.907~h (frame b, prewhitened data). Ten phase bins are used.}
\label{swkbkr}
\end{figure}

We can find the pulse shape of the oscillation by folding all the de-trended light-curves with a period of 3.771~h.  The folded light-curve is shown in Fig.~\ref{swkbkr}(a). The pulse shape of the oscillation is quasi-sinusoidal and slightly asymmetric. The prewhitened data folded with the 3.907~h period are presented in Fig.~\ref{swkbkr}(b) for comparison.

\begin{figure}
\includegraphics[width=84mm]{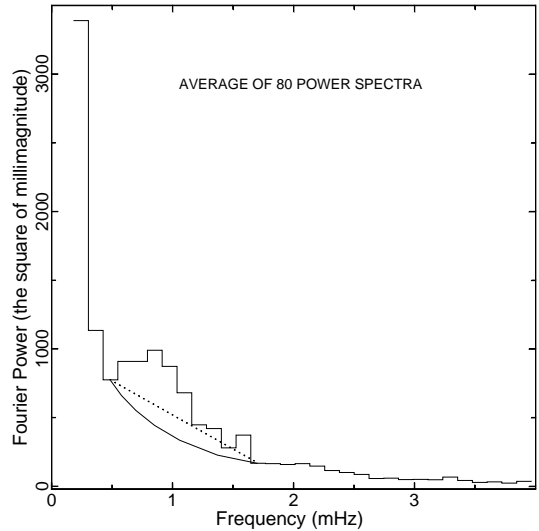}
\caption{Low frequency part of the average power spectrum of the KR~Aur data that are subdivided into segments. This power spectrum reveals the QPO hump at frequencies in the range 0.6--1.6~mHz. The dotted line and the arc indicate possible levels of the flickering below this QPO hump.}
\label{qpo}
\end{figure}

As mentioned, the light-curves of KR~Aur (Fig.~\ref{krblkr}) show the QPOs. When analysing QPOs, it is unnecessary to incorporate data into the common time series because OPQs are only weakly coherent. Instead we subdivided the de-trended light-curves into segments having a length of 137 min (1024 points) that are taken with a 50 per cent overlapping. This overlapping is advantageous because it increases the number of the power spectra used for averaging. Such an overlapping was widely used in the analysis of 1-3 second QPOs in polars (e.g. \citealt{larsson92}). This overlapping cannot change noise characteristics of the data because the data in the overlapping segments are the same data but slightly displaced in time.  The average Fourier power spectrum of all these data segments (Fig.~\ref{qpo}) shows the pronounced QPO hump at frequencies in the range 0.6--1.6~mHz. 

This power spectrum allows us to estimate the QPO power and compare it to the flickering power. However, we must know the flickering level below the QPO hump. We can consider two possible cases. In the first case we assume that the flickering level below the QPO hump is the line connecting the first and last points of this hump. This level seems overestimated. In the second case the flickering level is the arc connecting the first and last points of the QPO hump (as shown in Fig.~\ref{qpo}). This level seems more likely. In addition, we must know the noise of the data. At highest frequencies, when the flickering power becomes negligible, the power spectrum must flatten out due to the white noise in the data. Examining the high frequency part of the average power spectrum, we find that the flickering in KR~Aur is detectable up to a frequency of 47~mHz. The noise level at higher frequencies corresponds to a white noise of 19~mmag. This is close to the white noise component of the light-curves (15~mmag), which we evaluated directly (see section \ref{observations}).   Then, excluding this white noise level and integrating the power spectrum over frequencies higher than 0.2~mHz, we find that the relative power of the QPOs averages 19 and 35 per cent of the flickering power for the first and second cases, accordingly. This QPO power seems considerable. 

\section{DISCUSSION}

\begin{figure}
\includegraphics[width=84mm]{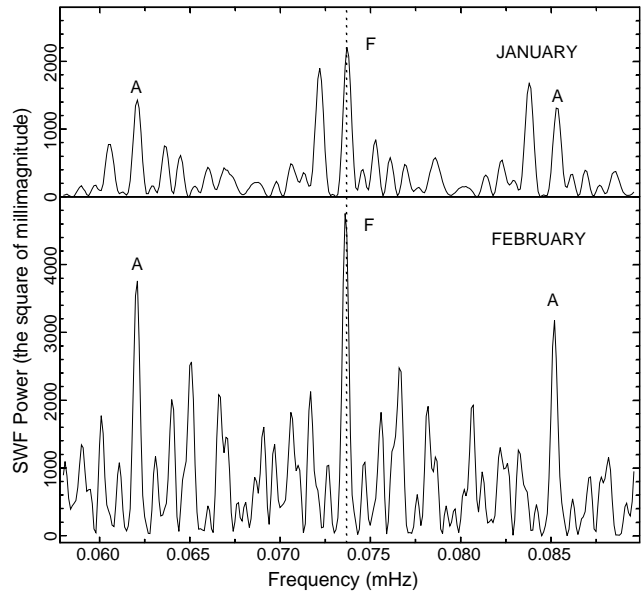}
\caption{SWF power spectra calculated for two groups of the data of KR~Aur.}
\label{partskr}
\end{figure}

\citet{shafter83} found the orbital period of KR~Aur equal to $3.9071 (\pm0.0007)$~h. The detected period is $3.771 (\pm0.005)$~h. This is 3.5 per cent less than $P_{orb}$. Such a small but clear discrepancy between the orbital and photometric periods is a feature of a permanent superhump system. Therefore this may indicate that the observed oscillation is a negative superhump. But brightness variations with periods slightly shorter than the orbital period are also typical of asynchronous polars (e.g. V1500~Cyg, \citealt{stockman88}). Unlike asynchronous polars, the superhumps usually show appreciable instability of their periods. Power spectra of our data divided into two groups hint at such an instability because the principal peaks are slightly displaced in frequency from each other (Fig.~\ref{partskr}). However, this displacement is small and can be caused by the noise of the data. None the less, we must exclude the possibility that the observed oscillation may be caused by a rotating magnetic white dwarf. The strong light polarization was never found in VY~Scl stars, and, therefore, they can contain only weakly magnetic white dwarfs similar to those of intermediate polars, rotation periods of which are much shorter than $P_{orb}$.

Since different kinds of superhumps are detected in CVs depending on their brightness states, it is important to know the brightness state of KR~Aur. During the main observations it was obvious that this star was in the bright state. Our 70-cm telescope do not allow us to observe stars fainter than 15~mag whereas KR~Aur was easily observable. As seen in Fig.~\ref{krblkr}, the average brightness of KR~Aur was decreasing. This trend might denote the transition to the low state. However, as mentioned, a month later we found the visible brightness of KR~Aur approximately at the same level with $B$=13.47 and $V$=13.35. In addition, we examined the light-curve of KR~Aur obtained by AAVSO observers (http://www.aavso.org/data/lcg) and also found that KR~Aur was in the bright state at least during four months after our observations. Thus, this trend does not denote the transition to the low state.

Unlike the positive superhumps which are usually explained by the prograde apsidal precession of the eccentric accretion disc, the negative superhumps can be explained by the retrograde precession of the line of nodes in an accretion disc tilted with respect to the orbital plane \citep{harvey95, skillman98, patterson02}. The disc tilt may be excited at the 3:1 Lindblad resonance due to gravitational perturbation of the secondary \citep{lubow92}. But this explanation may be applicable only to quiescent discs of short-period dwarf novae \citep{murray98}.  As a possible cause of the disc tilt, the influence of the magnetic field of the primary star was considered. However, \citet{murray02} found that the vertical structure of the disc induced by the magnetic field of the white dwarf co-rotates with the white dwarf and cannot generate the periods typical of negative superhumps. \citet{murray02} also investigated the influence of the magnetic field of the donor star. In this case the disc vertical structure may produce periods similar to observed in negative superhumps. However, \citet{foulkes05} found that the disc tilt induced by the donor star is transient in nature. Thus, plausible reasons for explanation of negative superhumps in nova-like variables are still not found.

The most well studied process for tilting the disc invokes the effect of radiation. This works well for luminous X-ray binaries, such as Her~X-1. The effect of radiation not only gives rise and maintains the disc tilt, but causes the disc to precess where the radiative precession can be prograde as well as retrograde \citep{wijers99, foulkes05}. But this requires very strong radiation from the central object, and is usually regarded not relevant for the white dwarf of modest luminosity in an ordinary CV \citep{patterson02}. However, supersoft X-ray binaries (SSBs) provide the unusual situation for accreting white dwarfs, in which most of the luminosity of the central source is generated by quasi-steady nuclear burning of the accreted material rather than by release of gravitational potential energy \citep{heuvel92}, and superorbital periods of brightness variations in some of SSBs can be explained by the effect of radiation \citep{murray98, southwell97}.

Although VY~Scl stars are not numerous and many of them are not well investigated, negative superhumps in such stars are frequently discovered. In the Ritter and Kolb catalogue (http://www.mpa-garching.mpg.de/RKcat) we find 23 VY~Scl stars, and six of them show negative superhumps. These are TT~Ari \citep{skillman98}, V751~Cyg \citep{patterson01}, DW~UMa, V442~Oph \citep{patterson02}, TW~Pis \citep{norton00}) and KR~Aur (this work). The large fraction of negative superhumpers among the VY~Scl stars can be related with their ability to undergo the extended low states which seem still enigmatic. Such a behavior is similar to the behavior of some SSBs, which also show optical high and low states. Indeed, \citet{greiner99} established that V751~Cyg was a luminous transient SSB during its optical low state when its unabsorbed bolometric luminosity was $ 6.5 \times 10^{36} (D/500 pc)^2 erg \,s^{-1} $. If nuclear burning is the correct interpretation of this X-ray flux of V751~Cyg during the optical low state, it may continue during the optical high state also although luminous supersoft X-ray emission in this state was not detected. If V751~Cyg behaves like some other SSBs, then in the optical high state it can predominantly emit in the UV band \citep{greiner99}. Therefore it seems possible that the negative superhumps in V751~Cyg may be caused by the effect of radiation. Although the estimated luminosity of V751~Cyg is somewhat lower than typically observed in classical SSBs, according to calculations made by \citet{foulkes05} it is sufficient for tilting the disc (see their table 2).

The basic properties of VY~Scl stars correspond surprisingly well to an extension of the SSB class, and the conjecture that all VY~Scl stars are SSBs seems viable \citep{greiner99}. Although luminous supersoft X-ray emission denoting nuclear burning   is directly observed in the only VY~Scl star V751~Cyg, some other VY~Scl stars also show indirect signs of nuclear burning. The strong UV radiation was found in the optical low state of TT~Ari, and as far back as in the early eighties it was suspected that this radiation might be caused by nuclear burning \citep{wagrau82}. The VY~Scl star BZ~Cam is surrounded by a faint emission nebula. Photo ionization by a canonical CV cannot account for the nebular excitation, and the emission line ratios of BZ~Cam's nebula and the nebula size are in agreement with the prediction of ionization by luminous $(10^{35}$--$10^{36} erg \,s^{-1})$ supersoft X-ray emission \citep{greiner01}. Furthermore, since we do not see other plausible reasons for negative superhumps in VY~Scl stars, as well as in luminous nova-like variables, we can consider negative superhumps as one more indirect sign of nuclear burning of the accreted material. 

\section{CONCLUSIONS}

\begin{enumerate}
\item We have found brightness variations with a period of 3.771~h in the optical light-curve obtained during 107 hours of observations of  the VY~Scl star KR~Aur in January and February 2004.
\item The semi-amplitude of these variations was observed approximately in the range 0.05--0.15~mag, showing changes from night to night.
\item The observed period is 3.5 per cent less than the orbital period, suggesting it is a negative superhump. 
\item Like other superhump systems, KR~Aur reveals QPOs on a time-scale of tens of minutes. The power of these QPOs averages 19--35 per cent of the flickering power at frequencies higher than 0.2~mHz. 
\item Distinctive features of VY~Scl stars and especially the discovery of luminous soft X-rays from V751~Cyg demonstrate that VY~Scl stars may contain white dwarfs, at which nuclear burning of the accreted material occur. Then the large fraction of negative superhumpers among VY~Scl stars suggests that negative superhumps in such stars may be caused by powerful radiation arising from nuclear burning.
\end{enumerate}


\begin{thebibliography}{}

\bibitem[\protect\citeauthoryear{Abbott \& Kleinman}{1994}]{abbott94}
Abbott T.M.C., Kleinman S.J., 1994, in Shafter A.W., ed, ASP Conf. Ser. Vol. 56, Interacting Binary Stars. Astron. Soc. Pac., San Francisco, p. 407

\bibitem[\protect\citeauthoryear{Bendat \& Piersol}{1986}]{bendat}
Bendat J.S., Piersol A.G., 1986, Random Data Analysis and Measurement
Procedures. John Wiley \& Sons, Ins., New York--Chichester--Brisbane--Toronto--Singapore

\bibitem[\protect\citeauthoryear{Deeg et al.}{1998}]{deeg98}
Deeg H.J. et al., 1998, A\&A, 338, 479

\bibitem[\protect\citeauthoryear{Foulkes, Haswell \& Murray}{Foulkes et al.}{2005}]{foulkes05}
Foulkes S.B., Haswell C.A., Murray J.R., 2005, MNRAS, 366, 1399

\bibitem[\protect\citeauthoryear{Greiner et al.}{1999}]{greiner99}
Greiner J., Tovmassian G.H., Di Stefano R., Prestwich A., Gonzalez-Riestra R., Szentasko L., Chavarria C., 1999, A\&A, 343, 183

\bibitem[\protect\citeauthoryear{Greiner et al.}{2001}]{greiner01}
Greiner J. et al., 2001, A\&A, 376, 1031

\bibitem[\protect\citeauthoryear{Harvey et al.}{1995}]{harvey95}
Harvey D., Skillman D.R., Patterson J., Ringwald F.A., 1995, PASP, 107, 551

\bibitem[\protect\citeauthoryear{Kato, Ishoka \& Uemura}{Kato et al.}{2002}]{kato02}
Kato T., Ishoka R., Uemura M., 2002, PASJ, 54, 1033

\bibitem[\protect\citeauthoryear{Kozhevnikov \& Zakharova}{2000}]{kozhevnikoviz}
Kozhevnikov V.P., Zakharova P.E., 2000, in Garzon F., Eiroa C., de Winter D., Mahoney T.J., eds, ASP Conf. Ser. Vol. 219, Disks, Planetesimals and  Planets. Astron. Soc. Pac., San Francisco, p. 381

\bibitem[\protect\citeauthoryear{Larsson}{1992}]{larsson92}
Larsson S., 1992, A\&A, 265, 133

\bibitem[\protect\citeauthoryear{Liller}{1980}]{liller80}
Liller M.H., 1980, AJ, 85, 1092


\bibitem[\protect\citeauthoryear{Lubow}{1992}]{lubow92}
Lubow S.H., 1992, ApJ, 398, 525

\bibitem[\protect\citeauthoryear{Murray \& Armitage}{1998}]{murray98}
Murray J.R., Armitage P.J., 1998, MNRAS, 300, 561

\bibitem[\protect\citeauthoryear{Murray et al.}{2002}]{murray02}
Murray J.R., Chakrabarty D., Graham A., Kramer W., Kramer L., 2002, MNRAS, 335, 247

\bibitem[\protect\citeauthoryear{Norton et al.}{2000}]{norton00}
Norton A.J., Beardmore A.P., Retter A., Buckley D.A.H., 2000, MNRAS, 312, 362

\bibitem[\protect\citeauthoryear{Patterson et al.}{1993}]{patterson93}
Patterson J., Thomas G., Skillman D.R., Diaz M., 1993, ApJS, 86, 235

\bibitem[\protect\citeauthoryear{Patterson et al.}{2001}]{patterson01}
Patterson J., Thorstensen J.R., Fried R., Skillman D.R., Cook L.M., Jensen L., 2001, PASP, 113, 72 

\bibitem[\protect\citeauthoryear{Patterson et al.}{2002}]{patterson02}
Patterson J. et al., 2002, PASP, 114, 1364

\bibitem[\protect\citeauthoryear{Schwarzenberg-Czerny}{1989}]{schwarzenberg}
Schwarzenberg-Czerny A., 1989, MNRAS, 241, 153

\bibitem[\protect\citeauthoryear{Shafter}{1983}]{shafter83}
Shafter A.W., 1983, ApJ, 267, 222

\bibitem[\protect\citeauthoryear{Singh et al.}{1993}]{singh93}
Singh J., Rao P.V., Agrawal P.C., Apparao K.M.V., Manchanda R.K., Sanwal B.B., Sarma M.B.K., 1993,
ApJ, 419, 337

\bibitem[\protect\citeauthoryear{Skillman et al.}{1998}]{skillman98}
Skillman D.R. et al., 1998, ApJ, 503, L67

\bibitem[\protect\citeauthoryear{Southwell, Livio \& Pringle}{Southwell et al.}{1997}]{southwell97}
Southwell K.A., Livio M., Pringle J.E., 1997, ApJ, 478, L29

\bibitem[\protect\citeauthoryear{Stockman, Schmidt \& Lamb}{Stockman et al.}{1988}]{stockman88}
Stockman H.S., Schmidt G.D., Lamb D.Q., 1988, ApJ, 332, 282

\bibitem[\protect\citeauthoryear{van den Heuvel et al.}{1992}]{heuvel92}
van den Heuvel E.P.J., Bhattacharya D., Nomoto K., Rappaport S.A., 1992, A\&A, 262, 97

\bibitem[\protect\citeauthoryear{Wagrau et al.}{1982}]{wagrau82}
Wagrau W., Drechsel H., Rahe J., Vogt N., 1982, A\&A, 110, 281

\bibitem[\protect\citeauthoryear{Wijers \& Pringle}{1999}]{wijers99}
Wijers R.A.M.J., Pringle J.E., 1999, MNRAS, 308, 207

\bibitem[\protect\citeauthoryear{Young et al.}{1991}]{young91}
Young A.T. et al., 1991, PASP, 103, 221

\end{thebibliography}
\end{document}